\documentclass[12pt]{article}
\usepackage{a4wide}
\usepackage{amsmath}
\usepackage{amssymb}
\usepackage{dsfont}
\usepackage[labelformat=simple]{subcaption}
\usepackage{graphicx}
\usepackage{algorithm}
\usepackage{algorithmicx}
\usepackage{algpseudocode}
\usepackage{booktabs}
\usepackage{varwidth}
\usepackage{url}
\usepackage{hyperref}
\usepackage[numbers,sort&compress]{natbib}
\usepackage[utf8x]{inputenc}

\numberwithin{equation}{section}
\numberwithin{figure}{section}
\numberwithin{table}{section}
\numberwithin{algorithm}{section}

\newcommand{\bracket}[1]{{\left\langle #1 \right\rangle}}
\newcommand{\In}{{\makebox[2.5em][l]{\textbf{in:}}}}
\newcommand{\Out}{{\makebox[2.5em][l]{\textbf{out:}}}}

\makeatletter
\newenvironment{breakablealgorithm}
  {
   \begin{center}
     \refstepcounter{algorithm}
     \hrule height.8pt depth0pt \kern2pt
     \renewcommand{\caption}[2][\relax]{
       {\raggedright\textbf{\ALG@name~\thealgorithm} ##2\par}%
       \ifx\relax##1\relax 
         \addcontentsline{loa}{algorithm}{\protect\numberline{\thealgorithm}##2}%
       \else 
         \addcontentsline{loa}{algorithm}{\protect\numberline{\thealgorithm}##1}%
       \fi
       \kern2pt\hrule\kern2pt
     }
  }{
     \kern2pt\hrule\relax
   \end{center}
  }
\makeatother

\algdef{SE}[VARIABLES]{Variables}{EndVariables}
   {\algorithmicvariables}
   {\algorithmicend\ \algorithmicvariables}
\algnewcommand{\algorithmicvariables}{\textbf{global variables}}

\title{\vskip-3cm{\baselineskip14pt
  \begin{flushleft}
      \normalsize TTK-17-24
  \end{flushleft}}
  \vskip1.5cm
  Mellin-Barnes meets Method of Brackets: A novel approach to Mellin-Barnes representations of Feynman integrals
}

\author{\small
  Mario Prausa
  \\[1em]
  {\small \it Institute for Theoretical Particle Physics and Cosmology}\\
  {\small \it RWTH Aachen University}\\
  {\small \it 52056 Aachen, Germany}\\[.5em]
  {\small \tt prausa@physik.rwth-aachen.de}
}  

\date{}

\begin{document}
  \maketitle
  \begin{abstract}
    In this paper, we present a new approach to the construction of Mellin-Barnes representations for Feynman integrals inspired by the Method of Brackets.
    The novel technique is helpful to lower the dimensionality of Mellin-Barnes representations in complicated cases, some examples are given.
  \end{abstract}

  \section{Introduction}
    The evaluation of multi-loop Feynman integrals is one of the basic building blocks in phenomenological and theoretical studies in quantum field theory.
    For this purpose, many techniques have been developed over the years.
    For an overview see e.g.~\cite{Smirnov:2012gma}.
    Among the most successful ones are the method of differential equations~\cite{Kotikov:1990kg,Kotikov:1991hm,Kotikov:1991pm}, Mellin-Barnes (MB) integral representations~\cite{Smirnov:1999gc,Tausk:1999vh}, and, for numerical evaluations, the method of sector decomposition~\cite{Binoth:2000ps,Binoth:2003ak,Binoth:2004jv}.

		After the construction of a MB representation for a given Feynman integral, one has a large amount of public tools at hand for their subsequent evaluation.
    For example one can resolve singularities~\cite{Czakon:2005rk,Smirnov:2009up}, expand in dimensional and analytic regulators~\cite{Czakon:2005rk}, perform an asymptotic expansion~\cite{mbasymptotics}, add up residues in terms of multi-fold sums~\cite{Ochman:2015fho} or numerically evaluate the integrals in the Euclidean domain~\cite{Czakon:2005rk}.
		For a long time the numerical evaluation of MB integrals with physical kinematics was an unresolved problem.
		However, significant progress was recently made also in that direction~\cite{Dubovyk:2016ocz,Dubovyk:2017cqw,Gluza:2016fwh} (see also \cite{Dubovyk:2016aqv} for a first application).

		Obviously, for all these applications one prefers to have a low number of MB integrations in the representation.
		This number strongly depends on the technique used to construct the representation.
    Two widely used techniques are the loop-by-loop approach~\cite{Gluza:2007rt,Gluza:2010rn} and the global approach~\cite{Blumlein:2014maa,Dubovyk:2016ocz}, both implemented in the public Mathematica package \texttt{AMBRE}~\cite{Gluza:2007rt,Gluza:2010rn,Blumlein:2014maa,Dubovyk:2016ocz}
    In the context of this paper, we denote a MB representation as better if it requires a lower number of MB integrations.

		Besides the already mentioned methods for the evaluation of Feynman integrals there exist also many less known techniques.
    One of them is the Method of Brackets~\cite{Gonzalez:2007ry,Gonzalez:2010,Gonzalez:2010uz}.
    This method is an improvement of an older technique called Negative Dimension Integration~\cite{Halliday:1987an}.

    The Method of Brackets defines a small set of simple rules which, when applied to a Schwinger parametrized Feynman integral, yields a set of multi-fold sums.
    Unfortunately, in many cases not all of these sums contribute to the final result and it is sometimes hard to tell which sum does contribute and which sum should be neglected.

    In this paper we modify the Method of Brackets so that it leads to a set of multi-dimensional MB integrals instead of a set of multi-fold sums.
    From this set of solutions a single multi-dimensional MB integral contains the full result of the Feynman integral.
		The ambiguity of the original method is therefore not present.
    
		Reference \cite{Gonzalez:2007ry} describes a factorization procedure for the Symanzik polynomials that appear in the Schwinger parametrization of Feynman integrals.
		This factorization reduces the multiplicity of the resulting multi-fold sums in the context of the original Method of Brackets.
		In our adapted version the same optimization helps to minimize the number of MB integrations in the constructed representation.
		This number is in some cases even smaller than for the best result the Mathematica package \texttt{AMBRE} can provide.
    The modified Method of Brackets is applicable for both planar and non-planar Feynman diagrams.

    In Section \ref{sect:brackets} we derive a set of rules for the adapted Method of Brackets in analogy to the rules defined in \cite{Gonzalez:2010uz}.
    Section \ref{sect:optimization} discusses the optimization of Symanzik polynomials.
    In Section \ref{sect:example} an example of the method is presented in great detail.
    At last, we compare our approach with the results of the \texttt{AMBRE} package for a couple of Feynman integrals in Section \ref{sect:ambre}.
  \section{The modified Method of Brackets} \label{sect:brackets}
    The original Method of Brackets is based on Ramanujan's master theorem~\cite{hardy1999ramanujan} which states that if a function $g(x)$ admits a Taylor expansion
    \begin{subequations} \label{eq:rmt}
      \begin{equation}
        g(x) 
        = 
        \sum\limits_{n=0}^\infty 
        G(n) 
        \frac{(-x)^n}{n!}
        \,,
      \end{equation}
      the integral over the parameter $x$ is given by
      \begin{equation}
        \int\limits_0^\infty dx\; 
        x^{\alpha-1} 
        g(x)
        =
        \Gamma(\alpha)
        G(-\alpha)\,. 
      \end{equation}
    \end{subequations}
    The similarity of this relation to the well-known Mellin-transform
    \begin{subequations} \label{eq:mellin}
      \begin{equation}
        f(x) 
        = 
        \int\limits_{c-i\infty}^{c+i\infty} 
        \frac{dz}{2\pi i}\; 
        x^z 
        F(z)
        \,,
      \end{equation}
      with
      \begin{equation}
        \int\limits_0^\infty dx\; 
        x^{\alpha-1} f(x)
        =
        F(-\alpha)
      \end{equation}
    \end{subequations}      
    allows reformulating the Method of Brackets in a way that leads to MB representations instead of multi-fold sums.

    Utilizing \eqref{eq:rmt}, the original Method of Brackets formulates a set of simple rules to rewrite a Schwinger parametrized Feynman integral \eqref{eq:schwinger} into a so-called presolution of the diagram - a multi-fold sum over $\Gamma$-functions and newly introduced objects called  brackets \cite{Gonzalez:2010uz}.
		The brackets in the presolution can then be eliminated using only linear algebra.

		In this section we present a similar set of rules, but our presolution will be a multi-dimensional MB integral instead of a multi-fold sum.
    
    \subsection{Schwinger parametrization} \label{sect:schwinger}
      The starting point to apply the Method of Brackets to Feynman integrals is Schwinger parametrization.
      An $L$-loop Feynman integral in Euclidean space-time is given by
      \begin{equation}
        I(a_1,\cdots,a_N) 
        = 
        \int\frac{d^dl_1}{\pi^{d/2}} 
        \cdots
        \int\frac{d^dl_L}{\pi^{d/2}} 
        \frac1{
          [P_1^2 + m_1^2]^{a_1}
          \cdots
          [P_N^2 + m_N^2]^{a_N}
        }\,,        
        \label{eq:feynman}
      \end{equation}
      where the momenta $P_i$ are linear combinations of loop momenta and external momenta.
      For physical Feynman integrals with a Minkowski space-time metric one can usually perform a Wick-rotation~\cite{Wick:1954eu} to transform the integral into the form \eqref{eq:feynman}.

      The Schwinger parameters $x_i$ are introduced for all propagators with the well known formula
      \[
        \frac1{[P_i^2 + m_i^2]^{a_i}}
        =
        \frac1{\Gamma(a_i)}
        \int\limits_0^\infty dx_i\;
        x_i^{a_i-1}
        e^{-x_i[P_i^2 + m_i^2]}\,.
      \]
      Afterwards the integrations over the loop-momenta can be performed loop-by-loop via
      \[
        \int\frac{d^dl}{\pi^{d/2}}
        e^{-\alpha l^2 + 2ql}
        =
        \alpha^{-d/2}
        e^{q^2/\alpha}\,.
      \]
      The result can be written as
      \begin{equation} \label{eq:schwinger}
        I(a_1,\cdots,a_N)
        =
        \frac1{\Gamma(a_1)\cdots\Gamma(a_N)}
        \int\limits_0^\infty dx_1\; x_1^{a_1-1}
        \cdots
        \int\limits_0^\infty dx_N\; x_N^{a_N-1}
        \;
        \frac{e^{-F/U - \sum_i x_i m_i^2}}{U^{d/2}}
        \,.
      \end{equation}
      The Symanzik polynomials $U$ and $F$ depend on the Schwinger parameters $x_i$ and can be read off directly from the Feynman graph.
      For an overview of the properties of these graph polynomials, see \cite{Bogner:2010kv}.

		\subsection{The Bracket}			
			The central object of the technique is the bracket, which is defined as
			\begin{equation} \label{eq:bracket}
        \bracket{\alpha}
        \equiv
			  \int\limits_0^\infty dx\; x^{\alpha-1}
        \,.
      \end{equation}
      Of course, this object by itself is not well-defined as the integral on the right-hand side is divergent for all $\alpha$.
      However, it makes sense inside a MB integral
      \begin{equation} \label{eq:bracketint}
        \int\limits_{c-i\infty}^{c+i\infty} \frac{dz}{2\pi i}
        \bracket{\alpha + z}
        F(z)
        =
        \int\limits_0^\infty dx\;
        \int\limits_{c-i\infty}^{c+i\infty} \frac{dz}{2\pi i}\;
        x^{\alpha + z - 1}
        F(z)
        =
        F(-\alpha)\,,
      \end{equation}
      where in the last step we used  equation \eqref{eq:mellin}.
      In contrast, the original Method of Brackets interprets this object inside a multi-fold sum using Ramanujan's master theorem \eqref{eq:rmt}.
    \subsection{The Rules} \label{sect:rules}
      The rules provided in this sub-section have to be applied successively to a Schwinger parameterized Feynman integral \eqref{eq:schwinger}.
      In doing so, rule~B has to be used multiple times if the Symanzik polynomials are given in optimized form (see Section~\ref{sect:optimization} for details).

      \subsubsection*{Rule A: Exponential functions}
        The exponential function in \eqref{eq:schwinger} is first split into factors using $e^{-\sum_i A_i} = \prod_i e^{-A_i}$ so that every exponent $A_i$ consists only of a monomial or a monomial divided by $U$.
        Afterwards the exponential functions are rewritten into contour integrals using the Cahen-Mellin formula
        \begin{equation} \label{eq:rule-A}
          e^{-A_i}
          =
          \int\limits_{c_i-i\infty}^{c_i+i\infty}
          \frac{dz_i}{2\pi i}
          A_i^{z_i} \,
          \Gamma(-z_i)\,.
        \end{equation}
        The contour is chosen such that all singularities coming from $\Gamma(-z_i)$ are to the right of the contour (i.e. $c_i < 0$).
        The validity of this equation can be checked by closing the contour at $|z_i|\rightarrow \infty$ to the right and using the residue theorem.

        The factor $A_i^{z_i}$ on the right-hand side of \eqref{eq:rule-A} should then be expanded to a product of powers, where the base is a single Schwinger parameter, the polynomial $U$, or one of the symbols introduced by the optimization procedure described in Section~\ref{sect:optimization}.
        After this, all powers of a common base have to be combined, e.g.
        \[
          U^{-d/2}
          \left(\frac{x_1 x_3}U \right)^{z_1}
          \left(\frac{x_1 x_4}U \right)^{z_2}
          =
          U^{-d/2-z_1-z_2} x_1^{z_1+z_2} x_3^{z_1} x_4^{z_2}
          \,.
        \]          

        This rule corresponds to rule~I in~\cite{Gonzalez:2010uz}.
      \subsubsection*{Rule B: Multinomials}
        Powers of multinomials occur after the insertion of the Symanzik polynomials $U$ or the re-substitution of the symbols introduced by the optimization procedure in Section~\ref{sect:optimization}.
        These powers can also be rewritten in terms of MB integrals using the formula
        \begin{equation} \label{eq:rule-B}
          \begin{split}
            &(A_1 + \cdots + A_J)^\alpha
            \\
            &\quad=
            \frac1{\Gamma(-\alpha)}
            \int\limits_{c_1-i\infty}^{c_1+i\infty}\frac{dz_1}{2\pi i}
            \cdots
            \int\limits_{c_J-i\infty}^{c_J+i\infty}\frac{dz_J}{2\pi i}
            \,
            \bracket{z_1 + \cdots + z_J - \alpha}
            \,
            A_1^{z_1}
            \cdots
            A_J^{z_J}
            \,
            \Gamma(-z_1)
            \cdots
            \Gamma(-z_J)\,.
          \end{split}          
        \end{equation}
        The formula can be derived by first applying Schwinger parametrization to the left-hand side of \eqref{eq:rule-B} and then using rule~A and the definition of the bracket \eqref{eq:bracket}.

        The factors $A_1^{z_1},\cdots,A_J^{z_J}$ on the right-hand side of \eqref{eq:rule-B} are treated in the same way as described for rule~A.

        This rule corresponds to rule~III in~\cite{Gonzalez:2010uz}.
      \subsubsection*{Rule C: Schwinger parameters}
        After the application of rule~A and B, the Schwinger integrals should all be of the form
        \[
          \int\limits_0^\infty dx_i\; 
          x_i^{L(a_1,\cdots; z_1,\cdots) - 1}
          \,,
        \]
        where $L(a_1,\cdots; z_1,\cdots)$ is a linear combination of the indices $a_j$ and the Mellin-Barnes variables $z_j$.
        These integrals can now be written as brackets using the definition \eqref{eq:bracket}:
        \[
          \int\limits_0^\infty dx_i\; 
          x_i^{L(a_1,\cdots; z_2,\cdots) - 1}
          =
          \bracket{L(a_1,\cdots;z_1,\cdots)}\,.
        \]          
        
        This rule corresponds to rule~II in~\cite{Gonzalez:2010uz}.
      \subsubsection*{Rule D: Eliminating the brackets}
        Applying the rules~A, B and C to a Schwinger parametrized Feynman integral results in a presolution of the form
        \[
          P
          =
          \int\limits_{c_1-i\infty}^{c_1+i\infty} \frac{dz_1}{2\pi i}
          \cdots
          \int\limits_{c_J-i\infty}^{c_J+i\infty} \frac{dz_J}{2\pi i}
          \,
          \bracket{\beta_1+\vec\alpha_1\cdot\vec z}
          \cdots
          \bracket{\beta_K+\vec\alpha_K\cdot\vec z}
          \,
          f(\vec{z})\,,
        \]
        where $J\geq K$ and $\vec{z} = (z_1,\cdots,z_J)^T$.

        We first consider the case $J=K$ and define a $K\times K$-matrix $A$ by
        \[
          A = \begin{pmatrix} \vec\alpha_1^T \\ \vdots \\ \vec\alpha_K^T \end{pmatrix}\,,
        \]
        where we assume for now its invertibility.
        A change of basis $\vec z = -A^{-1}\vec s\;$ leads to
        \begin{align*}
          P
          =
          \frac1{|\det A|}
          \int\limits_{d_1-i\infty}^{d_1+i\infty} \frac{ds_1}{2\pi i}
          \cdots
          \int\limits_{d_K-i\infty}^{d_K+i\infty} \frac{ds_K}{2\pi i}
          \,
          \bracket{\beta_1-s_1}
          \cdots
          \bracket{\beta_K-s_K}
          \,
          f(-A^{-1}\vec s)\,.
        \end{align*}
        Note the change in the integration contour to $(d_1,\cdots,d_K)^T = -A(c_1,\cdots,c_K)^T$.
        Now all MB integrations can be solved one by one using \eqref{eq:bracketint}:
        \[
          P
          =
          \frac1{|\det A|}
          f(-A^{-1} \vec\beta)\,,
        \]
        where $\vec\beta = (\beta_1,\cdots,\beta_K)^T$.

        In the case $J>K$, this formula can be used to solve $K$ out of the $J$ MB integrations.
        The result will be a $(J-K)$-dimensional MB integral.
        Without loss of generality we solve the MB integrals over $z_1,\cdots,z_K$ using the $K$ brackets while the $J-K$ integrals over $z_{K+1},\cdots,z_J$ should remain.
        Therefore, we arrange the first $K$ integration variables into a vector $\vec z_1 = (z_1,\cdots,z_K)^T$ and the variables of the remaining integrals into a vector $\vec z_2 = (z_{K+1},\cdots,z_J)^T$.
        Now, we can write down our last rule:
        \[
          \begin{split}
            &\int\limits_{c_1-i\infty}^{c_1+i\infty} \frac{dz_1}{2\pi i}
            \cdots
            \int\limits_{c_J-i\infty}^{c_J+i\infty} \frac{dz_J}{2\pi i}
            \,
            \bracket{\beta_1+\vec\alpha_1\cdot\vec z_1+\vec\gamma_1\cdot\vec z_2 }
            \cdots
            \bracket{\beta_K+\vec\alpha_K\cdot\vec z_1+\vec\gamma_K\cdot\vec z_2}
            \,
            f(\vec z_1,\vec z_2)
            \\
            &\quad=
            \frac1{|\det A|}
            \int\limits_{c_{K+1}-i\infty}^{c_{K+1}+i\infty} \frac{dz_{K+1}}{2\pi i}
            \cdots
            \int\limits_{c_J-i\infty}^{c_J+i\infty} \frac{dz_J}{2\pi i}
            f(-A^{-1} \vec\beta - A^{-1} C\,\vec z_2,\vec z_2)\,,
          \end{split}
        \]
        where the $K\times(J-K)$-matrix $C$ is given by $C = (\vec{\gamma_1}, \cdots, \vec{\gamma_K})^T$.
        The vector $\vec\beta$ and the matrix $A$ are again defined as $K$-dimensional quantities in the same way as before.

        The choice which integrals should be solved by this formula and which integrals should remain is somewhat arbitrary.
        There are $\binom JK$ possibilities.
        Some of them lead to a singular matrix $A$ and yield no solution.
        All other choices\footnote{Unfortunately, we cannot present a proof that a choice with $\det A \neq 0$ always exists.} lead to a possible MB representation for the full result, which implies that we only have to consider one of them.

        This is a major improvement from the original Method of Brackets, where the individual sum only gives a partial result for the Feynman integral and various choices (but not all) have to be considered to obtain a full result.

        We let the question unanswered, if some of the obtained MB representations are in some sense better than others.
        More studies are necessary to tackle this problem.

        This rule corresponds to rule~IV in~\cite{Gonzalez:2010uz}.
  \section{Optimization procedure} \label{sect:optimization}
    Rules~A to D applied to \eqref{eq:schwinger} are sufficient to obtain a MB representation for a given Feynman integral.
    However, the na\"ive application of these rules often leads to a huge number of MB integrations in the result.

    Better MB representations can be achieved by first analyzing the Symanzik polynomials $U$ and $F$ as well as the polynomial $\sum_i x_i m_i^2$ for sub-expressions (polynomials of Schwinger parameters) that appear multiple times.
    These common sub-expressions are then substituted by new variables which are treated as Schwinger parameters.
    This can be done recursively.

    The Schwinger parametrized Feynman integral \eqref{eq:schwinger} can now be treated with the set of rules given in Sub-Section~\ref{sect:rules} as before.
    However, after the first application of rule~B (to the power of base $U$), the intermediate result contains powers of the variables introduced by the optimization, such that rule~C cannot yet be applied.
    These powers have to be, after the corresponding sub-expressions are re-substituted, treated by rule~B as well.
    Only after all optimization variables have been eliminated, one can continue with rule~C. 

		This procedure reduces the number of MB integrals in the result significantly.
		If a polynomial $\xi$ with $N$ terms appears $J$ times in $U$, $F$ and $\sum_i x_i m_i^2$, the substitution of $\xi$ decreases the number of terms in these polynomials by $J(N-1)$.
		After rule~A and the first application of rule~B, the number of MB integrals is therefore reduced by $J(N-1)$ as well.
		However, the other application of rule~B, after $\xi$ is re-inserted, produces $N$ additional MB integrals and one additional bracket.
		In the end, this optimization leads therefore to a reduction of the number of MB integrals in the final result (after rule~D) by $(J-1)(N-1)$.

    This optimization approach was first proposed in \cite{Gonzalez:2007ry} in the context of the original Method of Brackets.
    An algorithm to find common sub-expressions in a list of polynomials is given in Appendix~\ref{sect:subex}.
  
  \section{Example} \label{sect:example}
    \begin{figure}
      \centering
      \includegraphics[scale=1.5]{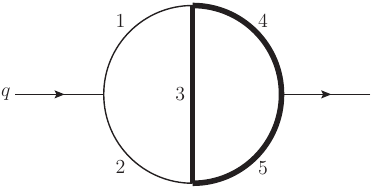}
      \caption{Two-loop propagator diagram. Bold (thin) lines represent massive (massless) propagators.}
      \label{fig:example}
    \end{figure}
    
    As an example, we consider the two-loop propagator diagram in fig.~\ref{fig:example}.
		The corresponding Feynman integral is given by
    \begin{align*}
      &I(a_1,\cdots,a_5)
      \\
      &\quad=
      \int\frac{d^dl_1}{\pi^{d/2}}
      \int\frac{d^dl_2}{\pi^{d/2}}
      \frac1{
        [l_1^2]^{a_1} 
        [(l_1-q)^2]^{a_2}
        [(l_1-l_2)^2 + m^2]^{a_3}
        [l_2^2 + m^2]^{a_4}
        [(l_2 - q)^2 + m^2]^{a_5}
      }
      \,,
    \end{align*}
    and the Schwinger parametrization by
    \[
      I(a_1,\cdots,a_5)
			=
			\frac{1}{\Gamma(a_1)\cdots\Gamma(a_5)}
			\int\limits_0^\infty dx_1\; x_1^{a_1-1}
			\cdots
			\int\limits_0^\infty dx_5\; x_5^{a_5-1}
      \;
			\frac{e^{-F/U - m^2(x_3+x_4+x_5)}}{U^{d/2}}
      \,,
    \]
    with
    \begin{align*}
      U &= x_2 x_5 + x_1 x_3 + x_1 x_5 + x_3 x_4 + x_1 x_4 + x_2 x_3 + x_3 x_5 + x_2 x_4 
      \,, \\
      F &= q^2 (x_1 x_2 x_5 + x_1 x_2 x_3 + x_1 x_3 x_5 + x_1 x_4 x_5 + x_1 x_2 x_4 + x_2 x_4 x_5 + x_2 x_3 x_4 + x_3 x_4 x_5) \,.
    \end{align*}
    A na\"ive application of rules~A to D without optimization would lead to a 13-fold MB representation.

    In order to reduce this number, we first identify common sub-expressions in $U$, $F$ and $\sum_i x_i m_i^2 = m^2(x_3+x_4+x_5)$  and replace them by new variables $r_i$:
    \begin{align*}
      r_1 &= x_3 + x_4\,, \\
      r_2 &= r_1 + x_5\,, \\
      r_3 &= r_2 x_1 + x_3 x_4\,, \\
      U &= r_2 x_2 + r_3 + x_3 x_5\,, \\
      F &= q^2 (x_2 x_4 x_5 + r_1 x_1 x_5 + r_3 x_2 + x_3 x_4 x_5)\,, \\
      \sum\limits_i x_i m_i^2 &= m^2 r_2\,.
    \end{align*}
    Rule~A then leads to 
    \begin{align*}
      I(a_1,\cdots,a_5)
			&=
			\int\limits_0^\infty dx_1
			\cdots
			\int\limits_0^\infty dx_5 
      \int\limits_{c_1-i\infty}^{c_1+i\infty} \frac{dz_1}{2\pi i}
      \cdots
      \int\limits_{c_5-i\infty}^{c_5+i\infty} \frac{dz_5}{2\pi i}
      \;
      (q^2)^{z_{1234}}
      (m^2)^{z_5}
      \frac{\Gamma(-z_1)\cdots\Gamma(-z_5)}{\Gamma(a_1)\cdots\Gamma(a_5)}
      \\ &\qquad
      \cdot
      U^{-d/2-z_{1234}}
      r_1^{z_2}
      r_2^{z_5}
      r_3^{z_3}
      x_1^{a_1+z_2-1}
      x_2^{a_2+z_{13}-1}
      x_3^{a_3+z_4-1}
      x_4^{a_4+z_{14}-1}
      x_5^{a_5+z_{124}-1}
      \,,
    \end{align*}
    where we introduced the notation $z_{ijk\cdots} = z_i + z_j + z_k + \cdots$ (and later also $a_{ijk\cdots} = a_i + a_j + a_k + \cdots$).
    Note that we have combined all powers of a common base.
    As a next step, the Symanzik polynomial $U$ in optimized form is re-inserted and rule~B applied:
    \begin{align*}
      I(a_1,\cdots,a_5)
			&=
			\int\limits_0^\infty dx_1
			\cdots
			\int\limits_0^\infty dx_5
      \int\limits_{c_1-i\infty}^{c_1+i\infty} \frac{dz_1}{2\pi i}
      \cdots
      \int\limits_{c_8-i\infty}^{c_8+i\infty} \frac{dz_8}{2\pi i}
      \;
      (q^2)^{z_{1234}}
      (m^2)^{z_5}
      \bracket{d/2+z_{1234678}} 
      \\ &\qquad
      \cdot
      \frac{\Gamma(-z_1)\cdots\Gamma(-z_8)}{\Gamma(a_1)\cdots\Gamma(a_5)\Gamma(d/2+z_{1234})}
      \\ &\qquad
      \cdot
      r_1^{z_2}
      r_2^{z_{56}}
      r_3^{z_{37}}
      x_1^{a_1+z_2-1}
      x_2^{a_2+z_{136}-1}
      x_3^{a_3+z_{48}-1}
      x_4^{a_4+z_{14}-1}
      x_5^{a_5+z_{1248}-1}
      \,,
    \end{align*}
    Now, rule~B must be used again three times for $r_3$, $r_2$ and $r_1$ in that order\footnote{The MB integration variables are sorted as in $z_1,\dots,z_9,z_a,\dots,z_e$.}:
    \begin{align*}
      I(a_1,\cdots,a_5)
			&=
			\int\limits_0^\infty dx_1
			\cdots
			\int\limits_0^\infty dx_5
      \int\limits_{c_1-i\infty}^{c_1+i\infty} \frac{dz_1}{2\pi i}
      \cdots
      \int\limits_{c_e-i\infty}^{c_e+i\infty} \frac{dz_e}{2\pi i}
      \;
      (q^2)^{z_{1234}}
      (m^2)^{z_5}
      \\ &\qquad
      \cdot
      \bracket{d/2+z_{1234678}} 
      \bracket{z_{9a} - z_{37}}
      \bracket{z_{bc} - z_{569}}
      \bracket{z_{de} - z_{2b}}
      \\ &\qquad
      \frac{\Gamma(-z_1)\cdots\Gamma(-z_e)}{\Gamma(a_1)\cdots\Gamma(a_5)\Gamma(d/2+z_{1234})\Gamma(-z_{37})\Gamma(-z_{569})\Gamma(-z_{2b})}
      \\ &\qquad
      \cdot
      x_1^{a_1+z_{29}-1}
      x_2^{a_2+z_{136}-1}
      x_3^{a_3+z_{48ad}-1}
      x_4^{a_4+z_{14ae}-1}
      x_5^{a_5+z_{1248c}-1}
      \,.
    \end{align*}
    Now we can apply rule~C to obtain the presolution
    \begin{align}
      I(a_1,\cdots,a_5)
			&=
      \int\limits_{c_1-i\infty}^{c_1+i\infty} \frac{dz_1}{2\pi i}
      \cdots
      \int\limits_{c_e-i\infty}^{c_e+i\infty} \frac{dz_e}{2\pi i}
      \,
      (q^2)^{z_{1234}}
      (m^2)^{z_5}
      \bracket{a_1+z_{29}}
      \bracket{a_2+z_{136}}
      \bracket{a_3+z_{48ad}}
      \;
      \nonumber \\ &\qquad
      \cdot
      \bracket{a_4+z_{14ae}}
      \bracket{a_5+z_{1248c}}
      \bracket{d/2+z_{1234678}} 
      \bracket{z_{9a} - z_{37}}
      \bracket{z_{bc} - z_{569}}
      \bracket{z_{de} - z_{2b}}
      \nonumber \\ &\qquad
      \cdot
      \frac{\Gamma(-z_1)\cdots\Gamma(-z_e)}{\Gamma(a_1)\cdots\Gamma(a_5)\Gamma(d/2+z_{1234})\Gamma(-z_{37})\Gamma(-z_{569})\Gamma(-z_{2b})}
      \,, \label{eq:presolution}
    \end{align}
    with 14 MB integrals and nine brackets, which leads to only \emph{five} MB integrations at the end.

    From the $\binom{14}{9} = 2002$ possibilities only 957 lead to a non-singular matrix $A$.
    We choose for the example the MB integrals over $z_1, z_2, z_3, z_4, z_7$ to remain.
    The vectors $\vec z_1$, $\vec z_2$ and $\vec\beta$ and the matrices $A$ and $C$ defined in rule~D read
    \begin{gather*}
      \vec z_1 
      =
      \begin{pmatrix} z_5 \\ z_6 \\ z_8 \\ z_9 \\ z_a \\ z_b \\ z_c \\ z_d \\ z_e \end{pmatrix}
      \,, \quad
      \vec z_2 
      =
      \begin{pmatrix} z_1 \\ z_2 \\ z_3 \\ z_4 \\ z_7 \end{pmatrix}
      \,, \quad
      \vec\beta
      =
      \begin{pmatrix} a_1 \\ a_2 \\ a_3 \\ a_4 \\ a_5 \\ \frac d2 \\ 0 \\ 0 \\ 0 \end{pmatrix}
      \,, \\[8pt]
      A 
      =
      \begin{pmatrix}
        0 & 0 & 0 & 1 & 0 & 0 & 0 & 0 & 0 \\
        0 & 1 & 0 & 0 & 0 & 0 & 0 & 0 & 0 \\
        0 & 0 & 1 & 0 & 1 & 0 & 0 & 1 & 0 \\
        0 & 0 & 0 & 0 & 1 & 0 & 0 & 0 & 1 \\
        0 & 0 & 1 & 0 & 0 & 0 & 1 & 0 & 0 \\
        0 & 1 & 1 & 0 & 0 & 0 & 0 & 0 & 0 \\
        0 & 0 & 0 & 1 & 1 & 0 & 0 & 0 & 0 \\
        -1 & -1 & 0 & -1 & 0 & 1 & 1 & 0 & 0 \\
        0 & 0 & 0 & 0 & 0 & -1 & 0 & 1 & 1
      \end{pmatrix}
      \,, \quad
      C 
      =
      \begin{pmatrix}
        0 & 1 & 0 & 0 & 0 \\
        1 & 0 & 1 & 0 & 0 \\
        0 & 0 & 0 & 1 & 0 \\
        1 & 0 & 0 & 1 & 0 \\
        1 & 1 & 0 & 1 & 0 \\
        1 & 1 & 1 & 1 & 1 \\
        0 & 0 & -1 & 0 & -1 \\
        0 & 0 & 0 & 0 & 0 \\
        0 & -1 & 0 & 0 & 0
      \end{pmatrix}        
      \,.
    \end{gather*}
    Using the formulas of rule~D, we have to substitute
    \begin{align*}
      z_5 &\rightarrow d - a_{12345} - z_{1234}\,, &
      z_6 &\rightarrow -a_2 - z_{13}\,, \\
      z_8 &\rightarrow - \frac d2 + a_2 - z_{247}\,, &
      z_9 &\rightarrow -a_1 - z_2, \\
      z_a &\rightarrow a_1 + z_{237}\,, &
      z_b &\rightarrow \frac d2 - 2a_1 - a_{234} - z_{147} - 2z_{23}\,, \\
      z_c &\rightarrow \frac d2 - a_{25} - z_{17}, &
      z_d &\rightarrow \frac d2 - a_{123} - z_3, \\
      z_e &\rightarrow -a_{14} - z_{12347}
    \end{align*}
    in \eqref{eq:presolution}, which yields the final five-dimensional MB representation
    \begin{align*}
      I(a_1,&\cdots,a_5) 
      \\
      &=
      \int\limits_{c_1-i\infty}^{c_1+i\infty} \frac{dz_1}{2\pi i}
      \cdots
      \int\limits_{c_4-i\infty}^{c_4+i\infty} \frac{dz_4}{2\pi i}
      \int\limits_{c_7-i\infty}^{c_7+i\infty} \frac{dz_7}{2\pi i}
      \;
      (m^2)^{d - a_{12345} - z_{1234}}
      (q^2)^{z_{1234}}
      \\ &\qquad
      \cdot
      \frac{
        \Gamma(-z_1)
        \cdots
        \Gamma(-z_4)
        \Gamma(-z_7)
        \Gamma(-d + a_{12345} + z_{1234})
        \Gamma(a_1 + z_2)
        \Gamma(-d/2 + a_{123} + z_3)
      }{
        \Gamma(a_1)
        \cdots
        \Gamma(a_5)
      }
      \\ &\qquad
      \cdot
      \frac{
        \Gamma(a_2 + z_{13})
        \Gamma(-d/2 + a_{25} + z_1 - z_7)
        \Gamma(-a_1 - z_{237})
        \Gamma(d/2 - a_2 + z_{247})
      }{
        \Gamma(d/2 + z_{1234})
        \Gamma(-d + 2a_{12} + a_{345} + 2z_{1234})
      }        
      \\ &\qquad
      \cdot
      \frac{
        \Gamma(a_{14} + z_{12347})
        \Gamma(-d/2 + 2a_1 + a_{234} + z_{147} + 2z_{23})
      }{
        \Gamma(-z_{37})
        \Gamma(- d/2 + 2a_1 + a_{234} + z_{1247} + 2z_3)
      }\,.
    \end{align*}
  \section{Comparison to \texttt{AMBRE}} \label{sect:ambre}
    \begin{figure}
      \centering
      \begin{subfigure}[b]{.19\textwidth}
        \centering
        \includegraphics[scale=.9]{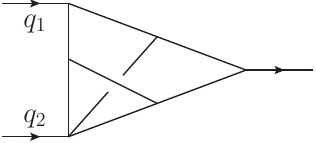}
        \caption{}
        \label{fig:int1}
      \end{subfigure}%
      \begin{subfigure}[b]{.19\textwidth}
        \centering
        \includegraphics[scale=.9]{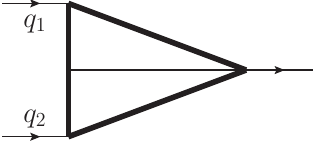}
        \caption{}
        \label{fig:int2}
      \end{subfigure}%
      \begin{subfigure}[b]{.19\textwidth}
        \centering
        \includegraphics[scale=.9]{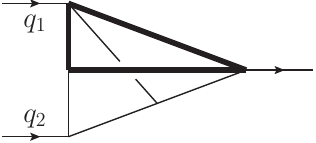}
        \caption{}
        \label{fig:int3}
      \end{subfigure}%
      \begin{subfigure}[b]{.19\textwidth}
        \centering
        \includegraphics[scale=.9]{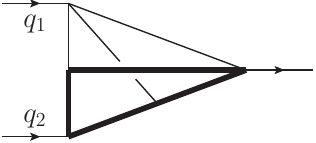}
        \caption{}
        \label{fig:int4}
      \end{subfigure}%
      \begin{subfigure}[b]{.19\textwidth}
        \centering
        \includegraphics[scale=.9]{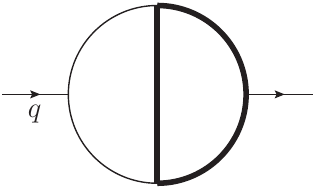}
        \caption{}
        \label{fig:int5}
      \end{subfigure}%
      \\
      \begin{subfigure}[b]{.19\textwidth}
        \centering
        \includegraphics[scale=.9]{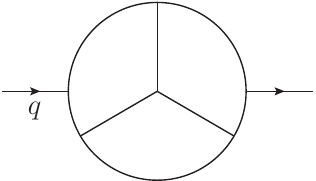}
        \caption{}
        \label{fig:int6}
      \end{subfigure}
      \begin{subfigure}[b]{.19\textwidth}
        \centering
        \includegraphics[scale=.9]{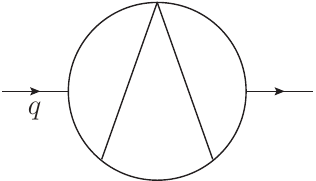}
        \caption{}
        \label{fig:int7}
      \end{subfigure}%
      \begin{subfigure}[b]{.19\textwidth}
        \centering
        \includegraphics[scale=.9]{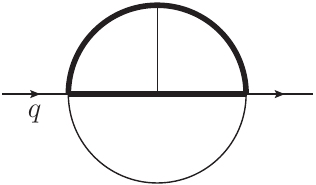}
        \caption{}
        \label{fig:int8}
      \end{subfigure}%
      \begin{subfigure}[b]{.19\textwidth}
        \centering
        \includegraphics[scale=.9]{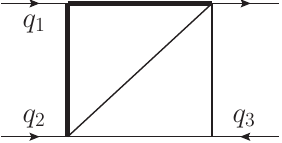}
        \caption{}
        \label{fig:int9}
      \end{subfigure}%
      \begin{subfigure}[b]{.19\textwidth}
        \centering
        \includegraphics[scale=.9]{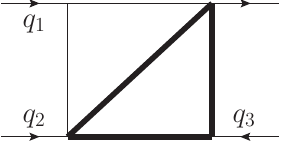}
        \caption{}
        \label{fig:int10}
      \end{subfigure} 
      \caption{ \label{fig:ambre}
        Example two- and three-loop diagrams. Bold (thin) lines represent massive (massless) propagators.
      }
    \end{figure}

    In this section we compare the MB representations of the diagrams in fig.~\ref{fig:ambre} obtained by our method to the representations constructed by the package \texttt{AMBRE}~\cite{Gluza:2007rt,Gluza:2010rn,Blumlein:2014maa,Dubovyk:2016ocz}.
    The kinematics for the triangle diagrams in fig.~\ref{fig:int1}--\subref{fig:int4} is
    \[
      q_1^2 = q_2^2 = 0\,, \quad
      q_1\cdot q_2 = \frac s2\,,
    \]
    for the propagator diagrams in fig.~\ref{fig:int5}--\subref{fig:int8}
    \[
      q^2 = s\,,
    \]      
    and for the two box diagrams in fig.~\ref{fig:int9} and \subref{fig:int10} 
    \[
      q_1^2 = q_2^2 = q_3^2 = 0\,, \quad
      q_1\cdot q_2 = \frac s2\,, \quad
      q_2\cdot q_3 = \frac t2\,, \quad
      q_1\cdot q_3 = -\frac s2 - \frac t2\,.
    \]
    The diagram in fig.~\ref{fig:int5} is the example from Section~\ref{sect:example}.

    For planar diagrams, we used the loop-by-loop approach implemented in \texttt{AMBRE} version 2.1~\cite{Gluza:2010rn} and tried out all permutations of the loop momenta to find the representation with a minimum number of MB integrations.
    We note that the quality of the loop-by-loop approach also depends on the momentum flow through the diagram\footnote{Thanks to Ievgen Dubovyk for pointing this out.}. 
    For non-planar diagrams, we used the global approach implemented in \texttt{AMBRE} version 3.1.1~\cite{Blumlein:2014maa,Dubovyk:2016ocz}.
    We tried to apply Barnes' first lemma to all representations, afterwards.
    Unfortunately, for none of the representations constructed by the Method of Brackets the lemma could be applied.

    The results are given in tab.~\ref{tab:ambre}.
    For the four diagrams in fig.~\ref{fig:int1}--\ref{fig:int4}, our approach leads to a lower-dimensional MB representation.
    However, for the four diagrams in fig.~\ref{fig:int5}--\subref{fig:int8} \texttt{AMBRE} is able to construct better results.
    For the diagrams in fig.~\ref{fig:int9} and fig.~\ref{fig:int10} both methods are comparable.

    As shown in tab.~\ref{tab:ambre}, our novel method can not provide a full replacement of the techniques implemented in \texttt{AMBRE} but could be helpful in some complicated cases. 
    \begin{table}
      \centering
      \begin{tabular}{@{}lccc@{}} \toprule
        diagram & Method of Brackets & \texttt{AMBRE} & planarity \\ \midrule
        fig.\ref{fig:int1}  & \textbf{7} & 13  & NP \\
        fig.\ref{fig:int2}  & \textbf{1} & 2 & P \\
        fig.\ref{fig:int3}  & \textbf{7} & 9  & NP \\
        fig.\ref{fig:int4}  & \textbf{7} & 8  & NP \\
        fig.\ref{fig:int5}  & 5 & \textbf{3}  & P \\
        fig.\ref{fig:int6}  & 9 & \textbf{4}  & P \\
        fig.\ref{fig:int7}  & 7 & \textbf{4} & P \\
        fig.\ref{fig:int8}  & 5 & \textbf{4} & P \\
        fig.\ref{fig:int9}  & \textbf{2} & \textbf{2}  & P \\
        fig.\ref{fig:int10} & \textbf{2} & \textbf{2}  & P \\
        \bottomrule
      \end{tabular}        
      \caption{The number of MB integrations of the representation constructed by the Method of Brackets compared to the best representation constructed by the \texttt{AMBRE}-package~\cite{Gluza:2007rt,Gluza:2010rn,Blumlein:2014maa,Dubovyk:2016ocz}. The smaller number is marked in bold. The last column gives the planarity of the diagram (P = planar, NP = non-planar).}
      \label{tab:ambre}
    \end{table}

    We checked the representations obtained by \texttt{AMBRE} and the Method of Bracket for numerical agreement using the Mathematica packages \texttt{MBresolve}~\cite{Smirnov:2009up} and \texttt{MB}~\cite{Czakon:2005rk}.
    The numerical integration was performed via the \texttt{MBintegrate} function of the package \texttt{MB} using the integration method \texttt{Cuhre}~\cite{Berntsen:1991a,Berntsen:1991b} implemented in the \texttt{Cuba}-library~\cite{Hahn:2004fe}.
    For the kinematic variables, we chose the Euclidean values $s=-1/2$, $t=-1/3$ and a mass $m = 1$ for the massive propagators.
    All propagator powers were set to one and the $\epsilon$-expansion was performed to next-to-leading order.
    The most complicated representation for a numerical integration was the seven dimensional representation for fig.~\ref{fig:int7} obtained by the Method of Brackets.
    Here, the maximum number of evaluation points had to be set to $2\cdot10^9$ and the runtime was about 16 hours on 16 CPU cores to achieve an agreement of the three most significant digits of the numerical values.
    However, more advanced integration algorithms may help to improve the accuracy reached on a reasonable time scale even for such high dimensional MB integrals.
    For recent developments in that direction, see~\cite{Dubovyk:2016ocz,Dubovyk:2017cqw}.
    All numerical values also agree with independent results of the sector decomposition implementation \texttt{FIESTA}~\cite{Smirnov:2015mct}.
  \section{Conclusion}
    In this article we presented a new technique to construct MB representations.
    The approach is based on a reformulation of the Method of Brackets.
    Our modified Method of Brackets yields not only one but many possible MB representations where every single one is a valid representation of the full Feynman integral. 
    This is a major improvement to the original Method of Brackets, where the question which solutions contribute to the full result was sometimes hard to answer.

    A crucial part of the method is the optimization procedure.
    Here, one has to analyze the graph polynomials for common sub-expressions.
    With this optimization, the method is able to produce low-dimensional MB representations.
    A simple algorithm for this purpose is given in appendix~\ref{sect:subex}.

    The presented method can easily be implemented in a computer code.

    Besides the practical applications, the reformulation of the Method of Brackets might help to deepen the understanding of the original Method of Brackets.
    It seems to be possible to relate the solutions of the original Method of Brackets in terms of multi-fold sums to the sums over residues of the MB representations obtained from our modified version.

  \section*{Acknowledgments}
    The author wants to thank Robert Harlander, Fabian Lange, and the \texttt{AMBRE} collaboration for useful discussions and comments on the manuscript, and especially for Ievgen Dubovyk's help in compiling tab.~\ref{tab:ambre}.

    This work was supported by BMBF contract 05H15PACC1.
    The computing resources were granted by RWTH Aachen University under project rwth0119.
    The Feynman diagrams in this article have been drawn with \texttt{JaxoDraw}~\cite{Binosi:2008ig} based on \texttt{Axodraw}~\cite{Vermaseren:1994je}.

  \appendix
  \section{Common Sub-Expressions} \label{sect:subex}
  In this appendix, we present a possible algorithm to find common sub-expressions in a given list of polynomials.
  The recursive algorithm shown in alg.~\ref{alg:cse} is far from being optimal but it proved nevertheless successful for all our tests.

  The main function of the algorithm is \textsc{commonBinomials} starting at line~\ref{algl:commonBinomials}. 
  The argument $P$ is an array of polynomials.
  Polynomials are in turn represented as arrays of terms (monomials).
  Arrays all start at index one.
  The function \textsc{commonBinomials} should be called with $p_0 = t_0 = 1$ and an empty set $r$.
  The best optimization found by the algorithm is returned in the global variables $\hat P$ and $\hat r$, where $\hat r$ is a set of rules which, when repeatedly applied to the array of polynomials $\hat P$, leads back to $P$.

  The quality of an optimization is measured by the rank $\rho$ calculated by the function \textsc{calcRank} starting at line~\ref{algl:calcRank}.
  The rank $\rho$ minus the number of propagators gives the number of MB integrals in the result.
  The goal of the algorithm is, therefore, to find an optimization, where $\rho$ is minimal.
  
  The first part (lines~\ref{algl:array-J-start} to \ref{algl:array-J-end}) of the algorithm fills an array $J$ with all possible optimizations which can be performed at the current level of the recursion.
  In this step we scan for binomials appearing in $P$ more than once.
  The actual scan starts at term $t_0$ in polynomial $p_0$.
  These arguments to the function \textsc{commonBinomials} are used to prevent scans of regions already completed at a lower recursion-level.
  Lines~\ref{algl:occ1} and \ref{algl:occ2} find all occurrences of the binomial $b$ in all polynomials in $P$.
  These occurrences are stored as triplets $(p,t_1,t_2)$ in the set $B$, where the first term of the binomial is found at $P[p,t_1]$ and the second term at $P[p,t_2]$, and $t_2 > t_1$.

  If the polynomials in $P$ do not have common binomials anymore, $J$ is empty at line~\ref{algl:J-empty}.
  In that case the optimization is complete.
  If the rank $\rho$ of this optimization is the lowest so far, the optimization is stored in the global variables.

  If $J$ is not empty, there are still common binomials in $P$.
  In principle, we could now try out all optimizations in $J$ one-by-one and then recursively call \textsc{commonBinomials}.
  Unfortunately, for large polynomials the algorithm would not terminate in a feasible time.
  For that reason, we only try out the first $N$ optimizations with the largest number of common binomials.
  For a finite $N < \infty$, it is therefore not guaranteed that the algorithm finds the best possible optimization.
  In most test cases even very small values of $N$, eg. 3 or 4, were sufficicient to find very good obtimizations in only a few seconds. 

  The lines~\ref{algl:early-start} and \ref{algl:early-end} cause an early exit, if the optimization at the current state is, even in the best-case scenario, not capable of producing a final optimization with a new minimum rank.

  The function \textsc{commonBinomials} only returns rules with \emph{binomials} on the right-hand side.
  In case, a symbol introduced by the algorithm does not appear in the returned list of optimized polynomials and only once on the right-hand side of one rule, it can be re-substituted without changing the rank of the optimization.
  This re-substitutions leads then to new rules where the right-hand sides have more than two terms.

  \begin{breakablealgorithm}
    \caption{Algorithm to find common binomials in a list of polynomials}
    \label{alg:cse}
    \begin{algorithmic}[1]
      \Variables
        \State $\hat\rho \gets \infty$
        \State $\hat P \gets ()$
        \State $\hat r \gets \{\}$
      \EndVariables
      \Statex        
      \Function{calcRank}{$P,r$} $\rightarrow \rho$ \label{algl:calcRank}
        \State \In
          \begin{varwidth}[t]{\linewidth}
            \makebox[1em][r]{$P$}: an array of polynomials \\
            \makebox[1em][r]{$r$}: a set of rules
          \end{varwidth}
        \State \Out \makebox[1em]{$\rho$}: an integer
        \State $\rho \gets |r| - 1$
        \For{$\textbf{all } p\in P$}
          \State $\rho \gets \rho + |p|$
        \EndFor
        \State \Return $\rho$
      \EndFunction
      \Statex
      \Function{replaceBinomials}{$P,B$} $\rightarrow (P',r,p_0,t_0)$
        \State \In
          \begin{varwidth}[t]{\linewidth}
            \makebox[1em][r]{$P$}: an array of polynomials \\
            \makebox[1em][r]{$B$}: a set of triplets $(p,t_1,t_2)$
          \end{varwidth}            
        \State \Out
          \begin{varwidth}[t]{\linewidth}
            \makebox[1em][r]{$P'$}: an array of polynomials \\
            \makebox[1em][r]{$r$}: a rule \\
            \makebox[1em][r]{$p_0$}: an index to a polynomial in $P'$ \\
            \makebox[1em][r]{$t_0$}: an index to a term in $P'[p_0]$
          \end{varwidth}            
        \State $(p_0,t_0) \gets \textit{undef}$
        \State $P' \gets \text{an array with $|P|$ empty elements}$
        \State $m_1 \gets P[B[1,1],B[1,2]]$
        \State $m_2 \gets P[B[1,1],B[1,3]]$
        \State $\displaystyle b \gets \frac{m_1 + m_2}{{\rm gcd}(m_1,m_2)}$
        \State $\xi \gets \text{a new symbol name}$
        \State $r \gets$ the rule ``$\xi \rightarrow b$''
        \State $D \gets \{\}$

        \For{$p \gets 1 \text{ to } |P|$}
          \For{$t \gets 1 \text{ to } |P[p]|$}
            \If{$(p,t) \not\in D$}
              \If{$\exists(p,t_1,t_2) \in B : t = t_1 \vee t = t_2$}
                \State $m_1 \gets P[p,t_1]$
                \State $m_2 \gets P[p,t_2]$

                \State append $\xi\cdot{\rm gcd}(m_1,m_2)$ to $P'[p]$
                \If{$(p_0,t_0) = \textit{undef}$\,}
                  \State $(p_0,t_0) \gets (p,|P'[p]|)$
                \EndIf

                \State $D \gets D \cup \{(p,t_1),(p,t_2)\}$
              \Else
                \State append $P[p,t]$ to $P'[p]$
                \State $D \gets D \cup \{(p,t)\}$
              \EndIf
            \EndIf
          \EndFor
        \EndFor
        \State \Return $(P',r,p_0,t_0)$
      \EndFunction
      \Statex
      \Procedure{commonBinomials}{$P,r,p_0,t_0,N$} \label{algl:commonBinomials}
        \State \In 
          \begin{varwidth}[t]{\linewidth}
            \makebox[1em][r]{$P$}: an array of polynomials \\
            \makebox[1em][r]{$r$}: a set of rules \\
            \makebox[1em][r]{$p_0$}: an index to a polynomial in $P$ \\
            \makebox[1em][r]{$t_0$}: an index to a term in $P[p_0]$ \\
            \makebox[1em][r]{$N$}: an integer
          \end{varwidth}            
        \State $D \gets \{\}$   \label{algl:array-J-start} 
        \State $J \gets ()$ 
        \For{$t \gets 1 \text{ to } t_0-1$}
          \If{$(p_0,t,t_0) \notin D$}
            \State $m_1 \gets P[p_0,t]$
            \State $m_2 \gets P[p_0,t_0]$

            \State $\displaystyle b \gets \frac{m_1 + m_2}{{\rm gcd}(m_1,m_2)}$

            \State $B \gets \text{set of all occurrences of binomial $b$ in all polynomials in $P$}$   \label{algl:occ1}

            \State $D \gets D\cup B$
            \If{$|B| > 1$}
              \State add $B$ to $J$ 
            \EndIf
          \EndIf
        \EndFor
        \For{$p_1 \gets p_0\text{ to } |P|$}
          \State $\displaystyle t_0' = \begin{cases} t_0 & \text{if } p_1 = p_0 \\ 1 & \text{else} \end{cases}$
          \For{$t_1\gets t_0' \text{ to } |P[p_1]|$}
            \For {$t_2\gets t_1+1 \text{ to } |P[p_1]|$}
              \If{$(p_1,t_1,t_2) \notin D$}
                \State $m_1 \gets P[p_1,t_1]$
                \State $m_2 \gets P[p_1,t_2]$

                \State $\displaystyle b \gets \frac{m_1 + m_2}{{\rm gcd}(m_1,m_2)}$

                \State $B \gets \text{set of all occurrences of binomial $b$ in all polynomials in $P$}$   \label{algl:occ2}

                \State $D \gets D\cup B$
                \If{$|B| > 1$}
                  \State add $B$ to $J$ 
                \EndIf
              \EndIf
            \EndFor
          \EndFor
        \EndFor     \label{algl:array-J-end}
        \State $\rho \gets \textsc{calcRank}(P,r)$
        \If{$|J| = 0$}  \label{algl:J-empty}
          \If{$\rho < \hat\rho$}
            \State $\hat\rho \gets \rho$
            \State $\hat P \gets P$
            \State $\hat r \gets r$
          \EndIf
        \Else
          \State $\delta \gets 0$         \label{algl:early-start}
          \For {$\textbf{all } j\in J$}
            \State $\delta \gets \delta + |j| - 1$
          \EndFor          
          \If {$\rho - \delta < \hat\rho$}     \label{algl:early-end}
            \State sort $J$ by the length of the elements. Longest element first.
            \If{$|J| > N$}    \label{algl:resize-start}
              \State resize $J$ to length $N$
            \EndIf            \label{algl:resize-end}
            \For{$\textbf{all } j\in J$}
              \State $(P',r',p_0',t_0') \gets \textsc{replaceBinomials}(P,j)$
              \State call $\textsc{commonBinomial}(P',r\cup\{r'\},p_0',t_0',N)$
            \EndFor
          \EndIf
        \EndIf 
      \EndProcedure
    \end{algorithmic}
  \end{breakablealgorithm}
  \bibliography{mbbrackets}
\end{document}